\documentclass[10pt,letterpaper]{article}
\usepackage{amsmath,amssymb}

\usepackage{changepage}

\usepackage[utf8x]{inputenc}

\usepackage{textcomp,marvosym}

\usepackage{cite}

\usepackage{nameref,hyperref}

\usepackage[right]{lineno}

\usepackage{microtype}
\DisableLigatures[f]{encoding = *, family = * }

\usepackage[table]{xcolor}

\usepackage{array}

\newcolumntype{+}{!{\vrule width 2pt}}

\newlength\savedwidth

\raggedright
\usepackage[aboveskip=1pt,labelfont=bf,labelsep=period,justification=raggedright,singlelinecheck=off]{caption}

\makeatletter
\renewcommand{\@biblabel}[1]{\quad#1.}
\makeatother

\date{}

\usepackage{lastpage,fancyhdr,graphicx}
\usepackage{epstopdf}
\usepackage{amsmath}
\usepackage{amssymb}
\usepackage{amsfonts}
\usepackage{bbm} % sudo apt-get install texlive-fonts-extra
\usepackage{bm}
\usepackage[english]{babel} % Specify a different language here - english by default
\usepackage{graphicx}

\newcommand\revision[1]{#1}

\begin{document}
\vspace*{0.2in}

\begin{flushleft}
{\Large
\textbf\newline{Computing with Networks of Nonlinear Mechanical Oscillators} % Please use "title case" (capitalize all terms in the title except conjunctions, prepositions, and articles).
}
\newline
\\
Jean C. Coulombe\textsuperscript{1},
Mark C. A. York\textsuperscript{1},
Julien Sylvestre\textsuperscript{1*}
\\
\bigskip
\textbf{1} Department of Mechanical Engineering, Université de Sherbrooke, Sherbrooke, Canada
\\
\bigskip

% Use the asterisk to denote corresponding authorship and provide email address in note below.
* julien.sylvestre@usherbrooke.ca
\end{flushleft}

\section*{Abstract}
As it is getting increasingly difficult to achieve gains in the density and power efficiency of microelectronic computing devices because of lithographic techniques reaching fundamental physical limits, new approaches are required to maximize the benefits of distributed sensors, micro-robots or smart materials.
Biologically-inspired devices, such as artificial neural networks, can process information with a high level of parallelism to efficiently solve difficult problems, even when implemented using conventional microelectronic technologies.
We describe a mechanical device, which operates in a manner similar to artificial neural networks, to solve efficiently two difficult benchmark problems (computing the parity of a bit stream, and classifying spoken words).
The device consists in a network of masses coupled by linear springs and attached to a substrate by non-linear springs, thus forming a network of anharmonic oscillators.
As the masses can directly couple to forces applied on the device, this approach combines sensing and computing functions in a single power-efficient device with compact dimensions.

%\linenumbers

\section*{Introduction}

Massively-parallel networks of simple units with elementary non-linear processing capabilities, such as artificial neural networks, have been used for years as efficient and robust computing systems.
In general, a network of $N$ elements is described by the state \revision{column} vector ${\bf x}(t) \in \mathbb{R}^N$.
The state vector evolves with time $t$ as a stimulus $u(t)$ is imposed on the network, according to the rich dynamics created by the interconnection of the elements in the network.
In a particularly simple network architecture called ``reservoir computing'' \cite{jaeger_harnessing_2004}, an output function
\begin{equation}
y(t) = {\bf w}^T {\bf x}(t)
\end{equation}
is formed using a weight vector ${\bf w} \in \mathbb{R}^N$.
As the network is fed a training signal $u^*(t)$, the weights are adjusted to minimize an error function between $y(t)$ and a target function $y^*(t) = F[u^*(t)]$.
$F$ is a complicated transformation of the input signal, which is used to represent the computing capabilities of the network.
During this training phase, the internal structure of the network is left untouched, and only the output weights are adjusted.

Such ``reservoir computers'' can approximate the transformation $F$ correctly when their dynamics obey the echo state property \cite{yildiz_re-visiting_2012}, in which case the states ${\bf x}(t)$ do not depend on the stimuli $u(t-\tau)$ for a time $\tau$ that is sufficiently long.
A reservoir computer with the echo state property will exhibit useful computing capabilities when the weights obtained during the training phase can be used to form an output $y(t)$ that is close to $F[u(t)]$, for a new stimulus $u(t)$ similar enough to $u^*(t)$.
It is actually observed in numerical experiments that reservoir computers perform well when the networks driven by the stimulus $u(t)$ operate as systems with complex dynamics \cite{bertschinger_real-time_2004}.
As such, reservoir computers are an efficient approach to exploit architectures such as recurrent neural networks, which are Turing equivalent \cite{kilian_dynamic_1996}, but which are difficult to train using conventional methods.
In practice, reservoir computers have been shown to be accurate and resource-efficient solutions to a number of challenging problems (e.g. ref. \cite{jaeger_harnessing_2004}, \cite{buteneers_automatic_2011}, \cite{triefenbach_phoneme_2010}).

A useful characteristic of reservoir computers is their fixed internal structure.
The network is generally constructed randomly with a few deterministic constraints such as limits on the spectral radius of the network connection matrix \cite{lukosevicius_reservoir_2009}, for instance.
Following the network construction, different weight vectors can be computed to enable the network to perform different tasks.
This fixed structure is especially attractive from the point of view of hardware implementation, as it does not require interconnections with dynamically adjustable strengths between the network elements, or other forms of network adaptability.
A number of hardware implementations of reservoir computers have been discussed, including analog electronics \cite{schurmann_edge_2004}, self-assembled atomic switches \cite{stieg_emergent_2012}, optoelectronic devices \cite{paquot_optoelectronic_2012} and photonic devices \cite{duport_all-optical_2012}.

This motivates the search for hardware elements that can be arranged in a network to form a dynamical system able to respond to an external stimulus, and to exhibit the echo state property.
We are especially interested in dynamical systems which can be stimulated directly by physical forces, such as accelerations or mechanical pressure.
\revision{In this communication, we verify numerically the hypothesis} that non-linear (anharmonic) mechanical oscillators coupled by linear springs can perform non-trivial computations.
This opens up the possibility for miniature, energy-efficient computers. 
As the mechanical elements are sensitive to physical forces, it also blurs the boundary between sensors and computers, with great opportunities in control and signal processing, for instance.

\revision{The objective of this study is to demonstrate that a network of non-linear mechanical oscillators can perform complex computations within the framework of reservoir computing.
We choose a specific form for the network, which is described in details below, and show that a single instance of such a network can efficiently solve two widely different computing problems.
This establishes the usability of networks of mechanical oscillators as general-purpose computing devices, which are able to efficiently process information from complex physical stimuli.
}

\begin{figure}[ht]\centering
\includegraphics[width=0.4\textwidth]{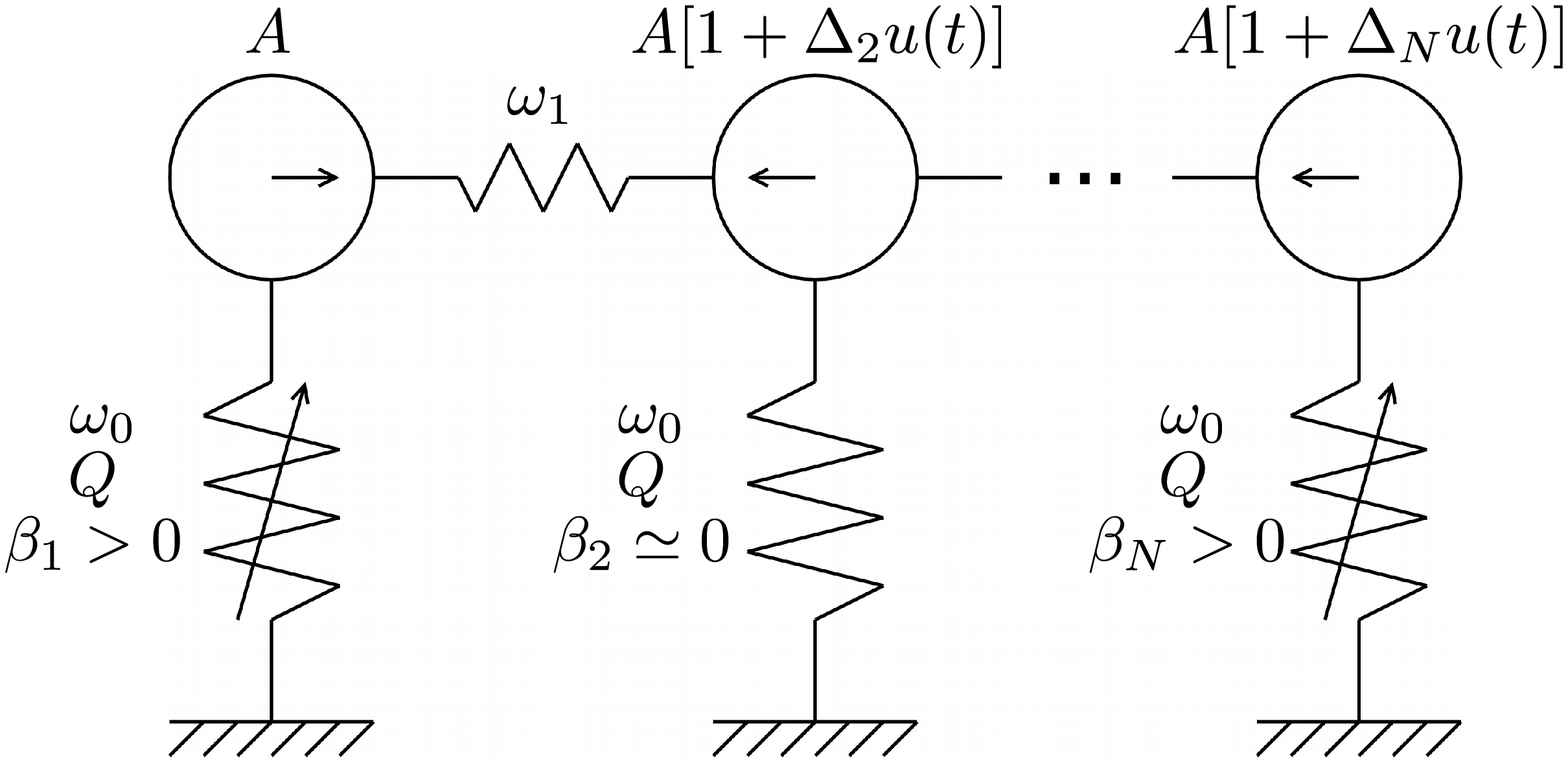}
\caption{
{\bf Schematic description of the computing network.}
The $N$ inertial masses (circles) arranged in a chain are coupled to neighbors by linear springs, and to a substrate by linear or non-linear springs, with damping.
A harmonic forcing, with amplitude possibly modulated by couplings to the input signal $u(t)$, is imposed on the masses.
}
\label{fig:schema}
\end{figure}

The following system is considered (see Fig \ref{fig:schema} for a schematic description):
\begin{eqnarray}
\ddot{x}_i(t) = -\frac{\omega_0}{Q} \dot{x}_i(t) - \omega_0^2 x_i(t) - \beta_i x_i^3(t) + \nonumber \\
A[1+\Delta_i u(t)]\cos(\Omega t) + \omega_1^2 [x_{i-1}(t) - 2 x_i(t) + x_{i+1}(t)], \label{eq:RC_Model}
\end{eqnarray}
where dots denote derivatives with respect to time, $i=1, ..., N$ and obvious modifications are made to the last term for $i=1$ \revision{($... + \omega_1^2[-x_1(t)+x_2(t)]$)} and $i=N$ \revision{($... + \omega_1^2[x_{N-1}(t)-x_N(t)]$)}.
Eq \ref{eq:RC_Model} describes a chain of nominally identical Duffing oscillators, except for the anharmonic term of strength $\beta_i$ which can vary in different oscillators.
Oscillators with $\beta_i = 0$ would be standard harmonic oscillators with resonance angular frequency $\omega_0 \sqrt{1-1/4Q^2}$ and quality factor $Q$.
The strength of the coupling between neighboring oscillators is parameterized by $\omega_1$.
% Note the definition of time scale in eq. \ref{eq:RC_Model} via the $\cos(t)$ driving term: if the oscillators were driven in a real application at a frequency $f$, then one time unit in the dimensionless time of eq. \ref{eq:RC_Model} corresponds to $1/2\pi f$ (seconds for $f$ in Hz).

The oscillators are driven by a harmonic term $\cos(\Omega t)$ at angular frequency $\Omega$, with a mean amplitude $A$ modulated by the input signal $u(t)$, scaled by the parameter $\Delta_i$.
When the drive amplitudes are large, the system can exhibit very complex dynamics, including extreme sensitivity to initial conditions (chaos).
At lower drive amplitudes, the dynamics can still be complex but are no longer chaotic.
Because of damping, the system exhibits the echo state property when its dynamics have a single attractor.
The existence of a single attractor (for a given class of input signals $u$) is verified numerically by obtaining a stable, high success probability when the oscillator network is trained for a particular task.

We present in section \ref{tasks} an example of an instance of a network described by Eq \ref{eq:RC_Model} (with $N=400$), which performs well on two significantly different benchmark computing tasks, namely the computation of non-trivial digital functions requiring memory (the parity function test), and the recognition of spoken digits.
This demonstrates that a given network (with fixed structure) can process information in widely different and complex ways.
This might be especially relevant technologically for space- and power-constraint applications, where it is beneficial to collocate a device sensing and processing functions.

%As we are using mechanical oscillators to create a reservoir computer, the input signal could be a force that couples directly to the oscillators.
%As an example, accelerations could be uniformly applied to all oscillators by inertial forces on their free masses, thus enabling the construction of accelerometers with built-in processing capabilities, for instance to generate a control signal for an actuator.
%Alternatively, the oscillators could be coupled electrostatically (at a frequency close to their resonance frequency) to a test mass moving at much lower frequencies under the influence of an external pressure (such as a sound wave). 
%This would enable the construction of microphones with built-in computing, for instance to identify spoken instructions.
%Similar applications would be most useful if the oscillator networks were small, light and power-efficient.

We consider as an example of a concrete implementation a network of thin doubly clamped silicon beams of length $l$ operated in their out-of-plane mode.
Such a network could be fabricated using conventional MEMS technologies.
The number density of oscillators would be approximately
% 1 µm x 0.2 µm = 0.2 µm² = 0.2 10^-6 mm²
\begin{equation}
5 \times 10^4\;{\rm mm}^{-2} \left(\frac{l}{10 \;\mu{\rm m}}\right)^{-2} \left(\frac{R_w}{0.1}\right)^{-1}, \nonumber
\end{equation}
where $R_w$ is the width-to-length ratio of the beams.
The resonance frequency of such oscillators is\cite{ekinci_nanoelectromechanical_2005}
\begin{equation}
\omega_0/2\pi = 8.5 \;{\rm MHz} \left(\frac{R_t}{0.01}\right) \left(\frac{l}{10 \;\mu{\rm m}}\right)^{-1}, \nonumber
\end{equation}
where $R_t$ is the thickness-to-length ratio of the beam.
As non-linear effects must be present in the oscillators for non-trivial computing capabilities to emerge, their oscillation amplitudes should be sufficiently large. 
The minimum amplitude for the onset of non-linear effects in a double clamped silicon beam is approximated by \cite{ekinci_nanoelectromechanical_2005}
\begin{equation}
x_{nl} = 20 \;{\rm nm} \left(\frac{Q}{100}\right)^{-1/2} \left(\frac{R_t}{0.01}\right) \left(\frac{l}{10 \;\mu{\rm m}}\right). \nonumber
\end{equation}
%where $R_t$ is the thickness-to-length ratio of the beam.
The energy in one oscillator is $m_{\rm eff} x_{nl}^2 \omega_0^2/2$, for $m_{\rm eff}$ the effective mass of the beam, so the mechanical power required to drive one oscillator in the reservoir is
\begin{equation}
0.5\;{\rm nW} \left(\frac{Q}{100}\right)^{-2} \left(\frac{R_t}{0.01}\right)^6 \left(\frac{R_w}{0.1}\right) \left(\frac{l}{10 \;\mu{\rm m}}\right)^{2}.  \nonumber
\end{equation}
%It is worth noting the exponent of 6 on $R_t$ in the power consumption expression above; membranes, such as graphene, could offer much lower power consumption levels than beams.
For comparison, reference \cite{merolla_million_2014} describes a state-of-the-art neuromorphic computing device implemented in the 28-nm CMOS technology with $10^6$ computing elements (artificial neurons).
This device achieves a neuron number density of $2\times 10^3\;{\rm mm}^{-2}$ (or $7\times 10^4\;{\rm mm}^{-2}$ when the peripheral circuitry not directly implementing the neurons is not considered).
Its power consumption per neuron is on the order of 90 nW.
As another example, reference \cite{yu_65k-neuron_2012} presents another neuromorphic chip (130-nm CMOS) with an artificial neuron number density of $3\times 10^3\;{\rm mm}^{-2}$ and power consumption per neuron of 4 nW.
The mechanical portion of the proposed reservoir computer with silicon beams could thus be smaller (by one order or magnitude) or more energy efficient (by one or two orders of magnitude) than a state-of-the-art microelectronic devices with the same number of computing element.
The density and power estimates of the oscillator network do not include the oscillator motion sensing, summation and amplification, which are expected to be implemented in efficient analog electronics, possibly using advanced packaging schemes such a heterogeneous integration.

Reservoir computers made of a network of coupled anharmonic mechanical oscillators therefore appear as interesting candidates for miniature, power-efficient devices that can be driven directly by physical signals (external fields, inertial or pressure forces, etc.), potentially allowing the creation of sensors with complex computing capabilities.
Numerical simulations demonstrating the computing capabilities of a mechanical oscillator network are presented in section \ref{tasks}, while the robustness of this computing model with respect to possible variability in the hardware implementation are discussed in section \ref{discussion}.

%------------------------------------------------

\section{Computing with a network of anharmonic oscillators} \label{tasks} 

The main result of this section is a numerical demonstration that a single instance of a network of coupled anharmonic oscillators, as described by Eq \ref{eq:RC_Model}, can perform well on different computing benchmarks.
All numerical results are obtained with the same oscillator network, excepted where explicitly mentioned for robustness evaluations.
It should be emphasized that the particular parameters and structure of the network presented below were only selected to demonstrate the usefulness of networks of oscillators as computers; the optimization of these parameters and structure to achieve better performances on specific tasks will be the subject of future communications.

The network consists of a long chain of $N=400$ oscillators, each with fundamental angular frequency $\omega_0=1.3$ and quality factor $Q=60$.
The oscillators are randomly assigned a strong ($\beta=1$) non-linearity with a probability of 25\%, and are otherwise assigned a weak ($\beta=0.005$) non-linearity (the same random allocation is used in all numerical experiments).
Neighboring oscillators in the chain are coupled by a linear spring of strength $\omega_1 = 1.5$.
In addition, the strength $\Delta_i$ of the coupling between the signal $u(t)$ and the oscillators is randomly set to a fixed value $\Delta^*$ (benchmark problem-dependent) for 50\% of the oscillators, and is zero otherwise.
The amplitude $A$ driving uniformly all the oscillators depends on the particular benchmark problem (see below).

Eq \ref{eq:RC_Model} is integrated numerically using a Runge-Kutta method.
In order to extract the envelopes of the rapidly-oscillating signals $x_i(t)$, these are multiplied by $\cos(\Omega t)$, decimated by a factor of 10, and passed through a seventh order low-pass Butterworth filter.
The envelope signals so-obtained, labeled $\chi_i(t)$, contain only the low-frequency amplitude variations in the oscillator positions.
They are used to form the output signal
\begin{equation}
y(t) = \sum_{i=1}^N w_i \chi_i(t),
\end{equation}
where the weights $w_i$ are computed from the data accumulated during the training period.
During the training period, a signal $u(t)$ is applied to the network, producing the data matrix $\bm{\Xi}$, with $[\bm{\Xi}]_{ij} = \chi_i(t_j)$ for discrete times $t_j$, $j=1...M$. 
The $N$-by-$N$ matrix $\bm{\Xi} \bm{\Xi}^T$ is inverted (with Tikhonov regularization) to obtain the weights from the vector of the target function evaluated at the same time steps.
It should be noted that the matrix $\bm{\Xi} \bm{\Xi}^T$ can be computed in real time by accumulating $N(N+1)/2$ values at each time step. 
The training period is then followed by a measurement period, where the weights are fixed, the signal $u(t)$ continues to be applied to the network, and the network's ability to reproduce the target function correctly is measured.

The parity function is considered as a first benchmark, as it requires both memory and non-trivial non-linear computational capabilities \cite{busing_connectivity_2010}.
For this task, $u(t)$ is a binary signal that can randomly switch between the two states $-1$ and $+1$ whenever $t$ is an integer multiple of a period $T$.
The $n^{\rm th}$-order parity of $u$ is defined as
\begin{equation}
P_n(t) = \prod_{i=1}^n u(t-iT),
\end{equation}
and requires data between $t-nT$ and $t-T$ to be continuously computed.
Numerical experiments were performed with $T=65$ and $\Omega=1.14$, so that the input signal $u(t)$ was switching at most every $9.07$ cycles of the $\cos(\Omega t)$ drive.
The amplitude parameters were set to $A=0.8$ and $\Delta^*=0.7$. %%% NOTE: Acode * 0.5 = A
In order to increase the robustness of the estimation of the parity functions, the weights were computed on ten contiguous, equal-length sub-sections of the full oscillator chain during a training phase of duration $359T$.
Each sub-section had its own set of weights and was sufficiently long (40 oscillators) to produce a good estimate of the parity function for most inputs.
The ten estimates from the ten sub-sections were averaged, the sum was integrated over each period $T$, and the sign of the integral was used to decide between a value of -1 or of +1 for the parity function estimated by the network.

\begin{figure*}[ht]\centering % Using \begin{figure*} makes the figure take up the entire width of the page
\includegraphics[width=0.9\textwidth]{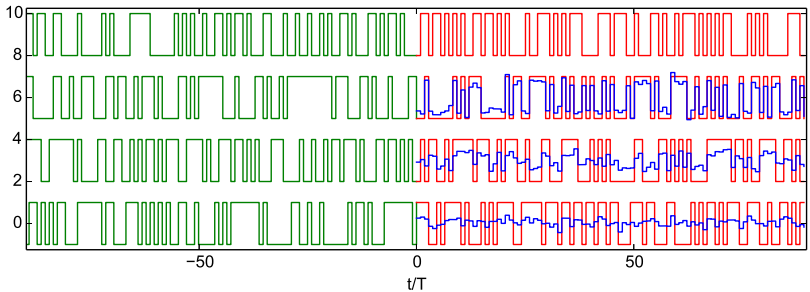} %fplotVote.eps}
\caption{
{\bf Computing parity functions with a network of coupled anharmonic oscillators.} 
Driving term $u(t)$, $P_3(t)$, $P_4(t)$, $P_5(t)$ (top to bottom, shifted vertically for clarity). Green curves correspond to the training phase ($t<0$). For $t>0$, red curves correspond to the target functions, while blue curves correspond to the network outputs. 
}
\label{fig:computing}
\end{figure*}

Fig \ref{fig:computing} shows a numerical example obtained for parity functions of order 3, 4 and 5.
Estimated parity values of +1 or -1 indicate that all ten sub-sections of the full chain, which are mostly equivalent to shorter independent chains with $N=40$, were all able to obtain the correct value.
On the other hand, estimated parity values around 0 (as observed more frequently for $P_5$) indicate that the equivalent shorter chains were unable to agree on a valid parity value.
As in Fig \ref{fig:computing}, it is observed in larger scale experiments that the accuracy of the network decreases with increasing order of the parity function, with a proportion of correctly estimated parity values of $100\%$ for $P_3$, $(93.48 \pm 0.0031)\%$ for $P_4$ and $(68.78 \pm 0.0058)\%$ for $P_5$. 
The training process appears to be relatively robust, with the $10^{\rm th}$ percentile of the proportion of correctly estimated parity values for repeated training runs (which differ only in the randomly generated training data) estimated at $86.2\%$ and $60.4\%$ for $P_4$ and $P_5$, respectively.

A convenient way to compare the capacitites of the oscillator networks to other results in the reservoir computing literature is to estimate their so-called memory capacity, as introduced in reference \cite{bertschinger_real-time_2004}.
As bits are distributed equally between $-1$ and $+1$ in both the input and output of the parity benchmark, the mutual information\cite{bertschinger_real-time_2004} is estimated using
\begin{equation}
MI_{\tau} = p_{\tau}\log_2(2p_{\tau}) + (1-p_{\tau})\log_2(2(1-p_{\tau})),
\end{equation}
for $p_{\tau}$ the success probability of the delayed $3^{\rm rd}$ order parity function, defined as
\begin{equation}
\prod_{i=0}^2 u(t-(i+\tau)T).
\end{equation}
The memory capacity is then given by
\begin{equation}
\sum_{\tau=0}^4 MI_{\tau},
\end{equation}
with the sum truncated at $\tau=4$ because larger delays have negligible mutual information.
The mutual information for the delayed $P_3$ function is 1 bit for a delay $\tau \leq 2T$, and then drops rapidly to 0.44 bit for $\tau = 3T$ and less than 0.04 bit for $\tau \geq 4T$.
The resulting memory capacity is approximately 3.5 bits, similar to performance levels for large networks published in the reservoir computing literature (e.g. reference \cite{bertschinger_real-time_2004} presents a reservoir with 250 neurons with a memory capacity of 4.8 bits).

As another benchmark, we consider the classification of recorded time series for the spoken words ``zero'' to ``nine''.
This is a conventional benchmark for non-trivial classification tasks (e.g. ref. \cite{appeltant_information_2011}), when the NIST TI-46 data set \cite{liberman_mark_ti_1993} is used.
We use from this data set utterances of the words from 7 different female speakers. %% NOTE: REDO WITH FEMALE and MALE
In this task, the driving signal $u(t)$ is formed by concatenating time series of the utterances, padded with periods of silence (duration $70$). % PPer/2
The mean of each time series is removed, the time series is normalized by its standard deviation, and its absolute value is used for $u(t)$.
The time series are provided in TI-46 at a sampling rate of 12.5 kHz.
In order to adapt the time series to the dynamics of the simulated oscillators, we stretch every second of time series data to 97.2 time units in the numerical simulation, which is the equivalent of having oscillators driven at a frequency of 808 Hz ($\Omega=1$).
% 12500 samples (1s) --> Pper*12500/18000 = 97.2 (Pper=140)
% drive period = 2pi --> 97.2/2pi = 12500/# cycles --> f = 808 Hz 
As a result, the network mostly uses the low frequency content of the sound recordings to classify the utterances.
%While simulation computing times limit the amount of stretching (and therefore the equivalent oscillators drive frequency), physical implementations could very well operate at much higher frequencies.
%Whether of not this might create additional optimization opportunities relative to the results presented below will be the subject of future communications.
The amplitude parameters were set to $A=2$ and $\Delta^*=6$. %%% NOTE: Acode * 0.5 = A

The training was performed on 800 utterances chosen randomly between the ten digits.
One set of weights was computed for each digit.
As for the parity benchmark, the chain of $N=400$ oscillators was split into sub-sections to improve robustness, this time into 19 sub-sections of length 40, each overlapping the next by 20 oscillators.
Each sub-section had its own set of weights and produced a value that should be one if the digit corresponding to this set was spoken, and zero otherwise.
The values produced from the weights were integrated over each period where an utterance was submitted to the network.
As a result, for each utterance, the nineteen sub-sections for each of the ten digits produced a total of 190 numbers $c_{ij}$, with $i=1, ..., 10$ and $j=1, ..., 19$.
The classification from the network was then obtained using $10\times 9/2$ pair comparisons in a one-vs.-one manner, according to
\begin{equation}
\bm{D}^T_{ii'}(\bm{c}_{i} - \bm{c}_{i'}) \underset{i'}{\overset{i}{\gtrless}} T_{ii'}, \label{eq:binary}
\end{equation}
where $T_{ii'}$ is a threshold number, $\bm{D}_{ii'}$ is a vector of coefficients to linearly combine the components of the difference of vectors $\bm{c}_i$ and $\bm{c}_{i'}$, which correspond to $[\bm{c}_i]_j = c_{ij}$.
The digit $i$ or $i'$ that was returned earned one vote for each comparison, and the digit with the largest number of votes was returned by the network for the utterance.
The coefficient vectors were computed using Fisher's linear discriminant, according to
\begin{equation}
\bm{D}_{ii'} = (N_i \bm{\Sigma}_i + N_{i'} \bm{\Sigma}_{i'} + \lambda \bm{\mathbbm{1}})^{-1} (\bm{\mu}_i - \bm{\mu}_{i'}),
\end{equation}
where $N_i$ is the number of $i$ digits in the training data, $\bm{\Sigma}_i$ is the covariance matrix of vector $\bm{c}_i$, $\lambda$ is a small regularization parameter, $\bm{\mu}_i$ is the average of the vector $\bm{c}_i$, and similarly for $i'$.
Each threshold $T_{ii'}$ was adjusted to maximize the probability over the training set of Eq \ref{eq:binary} to return the correct digit when $i$ and $i'$ occurred with equal probability.

\begin{figure*}[ht]\centering % Using \begin{figure*} makes the figure take up the entire width of the page
\includegraphics[width=0.9\textwidth]{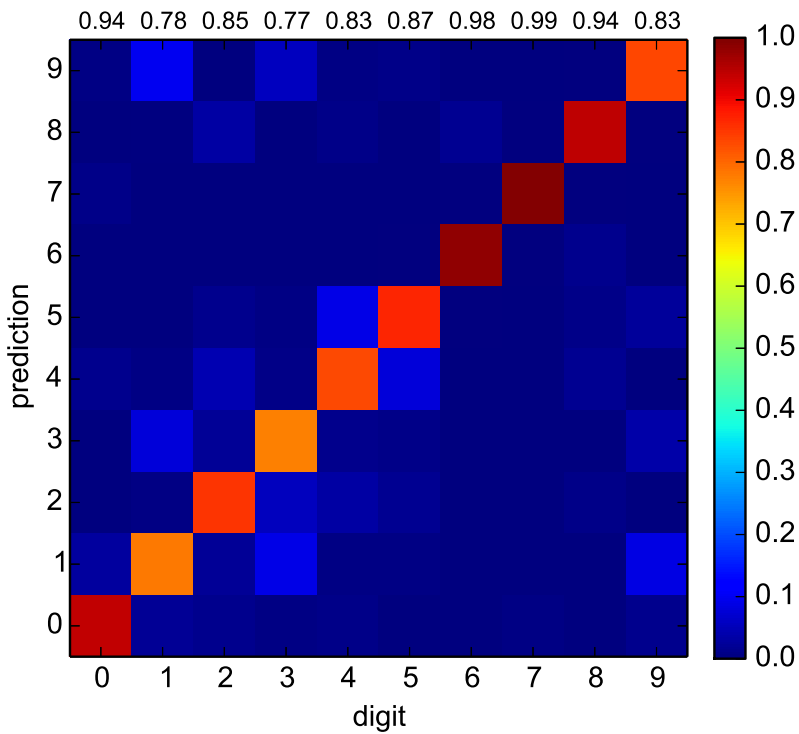} %Rerun_Call_w0_CALL.eps}
\caption{
{\bf Words classification benchmark results.} The color-scale indicates the probability that a digit presented to the device (columns) is classified to a certain value (lines) by the oscillators network. The numbers at the top of each column indicate the success probability (prediction matches the actual digit), estimated with an uncertainty of $\pm 1\%$ (95\% confidence level).
}
\label{fig:speech}
\end{figure*}

Fig \ref{fig:speech} presents results for the words classification benchmark.
The results correspond to the average performance of 25 different training runs, performed on the same network.
For all training runs, the variability in success rate is consistent with uncertainties from the finite sample size, indicating that the training procedure is repeatable.
The results are relatively good, with the network correctly classifying randomly chosen utterances in $(0.802 \pm 0.009)\%$ of the trials.
It can be seen in Fig \ref{fig:speech} that classification errors are principally made between small groups of digits such as $\{1, 3, 9\}$ and $\{4, 5\}$, indicating that it is harder for the network to discriminate between digits within these groups than between these and other digits.

The results can be compared to other experiments in the reservoir computing literature.
While success rates above 99\% have been reported (for five speakers instead of seven as in this work) for networks of 200 nodes (e.g. ref. \cite{rodan_minimum_2011, paquot_optoelectronic_2012}), these experiments all make use of elaborated pre-processing schemes (specifically, the Lyon cochlear ear model \cite{lyon_computational_1982}).
It has been shown in ref. \cite{verstraeten_isolated_2005} that pre-processing greatly affects the efficiency of spoken words classification, with some pre-processors (e.g. using Mel-frequency cepstrum coefficients) performing well under 70\% success rates for networks of 200 nodes.
The results presented here with the oscillator network do not include any pre-processing, except for the rectification of the input time series to form the input signal $u(t)$, and the normalization of the time series amplitudes.
This is intended to reflect a simple system where the sound pressure directly modulates the driving force on the oscillators (e.g. by displacing a membrane).
More complicated systems, for instance with modulation amplitudes for different oscillator groups that depend on the sound frequency, could in principle be significantly more efficient.

\section{Discussion} \label{discussion}
\begin{figure*}[ht]\centering % Using \begin{figure*} makes the figure take up the entire width of the page
\includegraphics[width=0.3\textwidth]{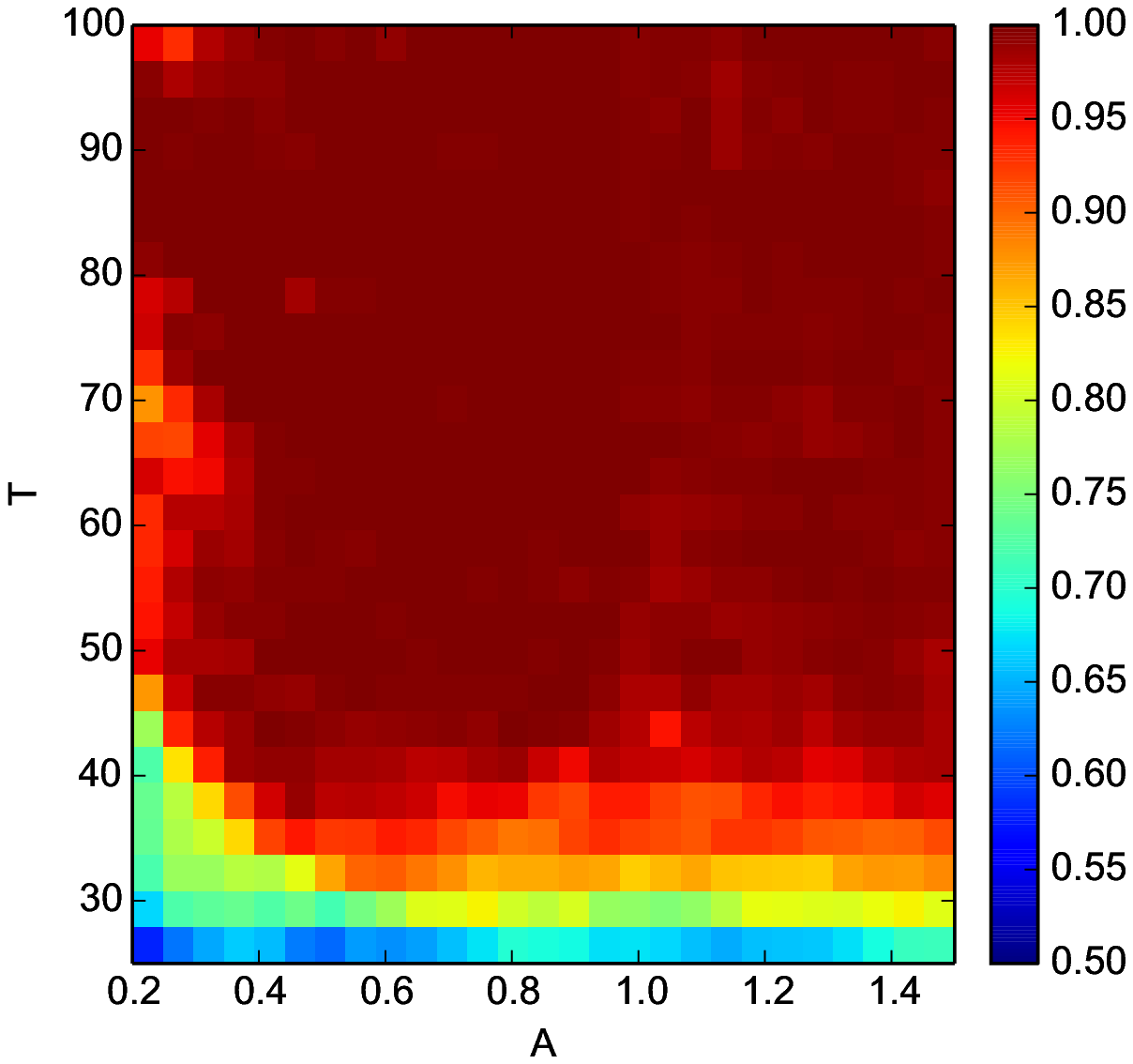} %Parity_P3.eps}
\includegraphics[width=0.3\textwidth]{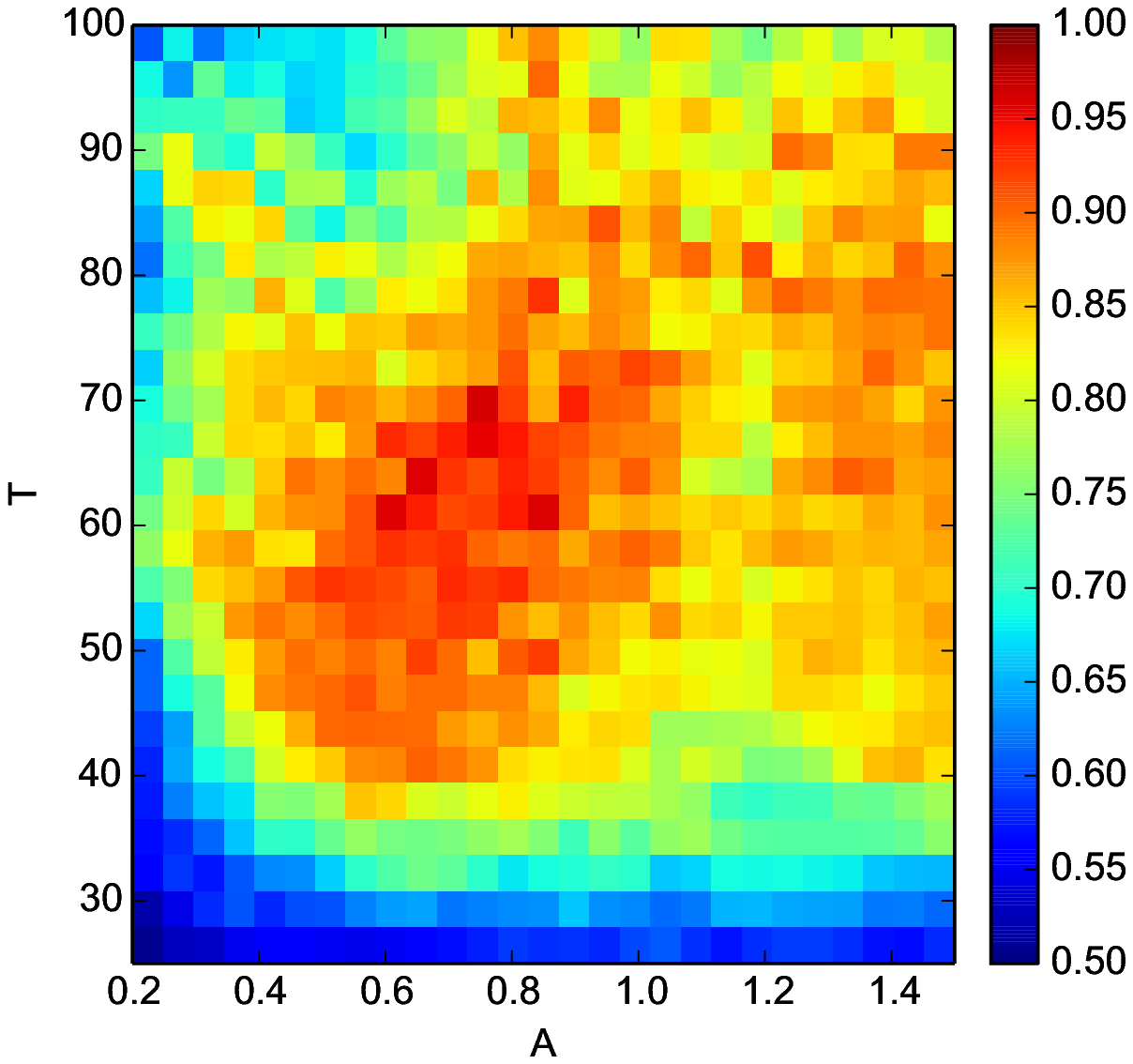} %Parity_P4.eps}
\includegraphics[width=0.3\textwidth]{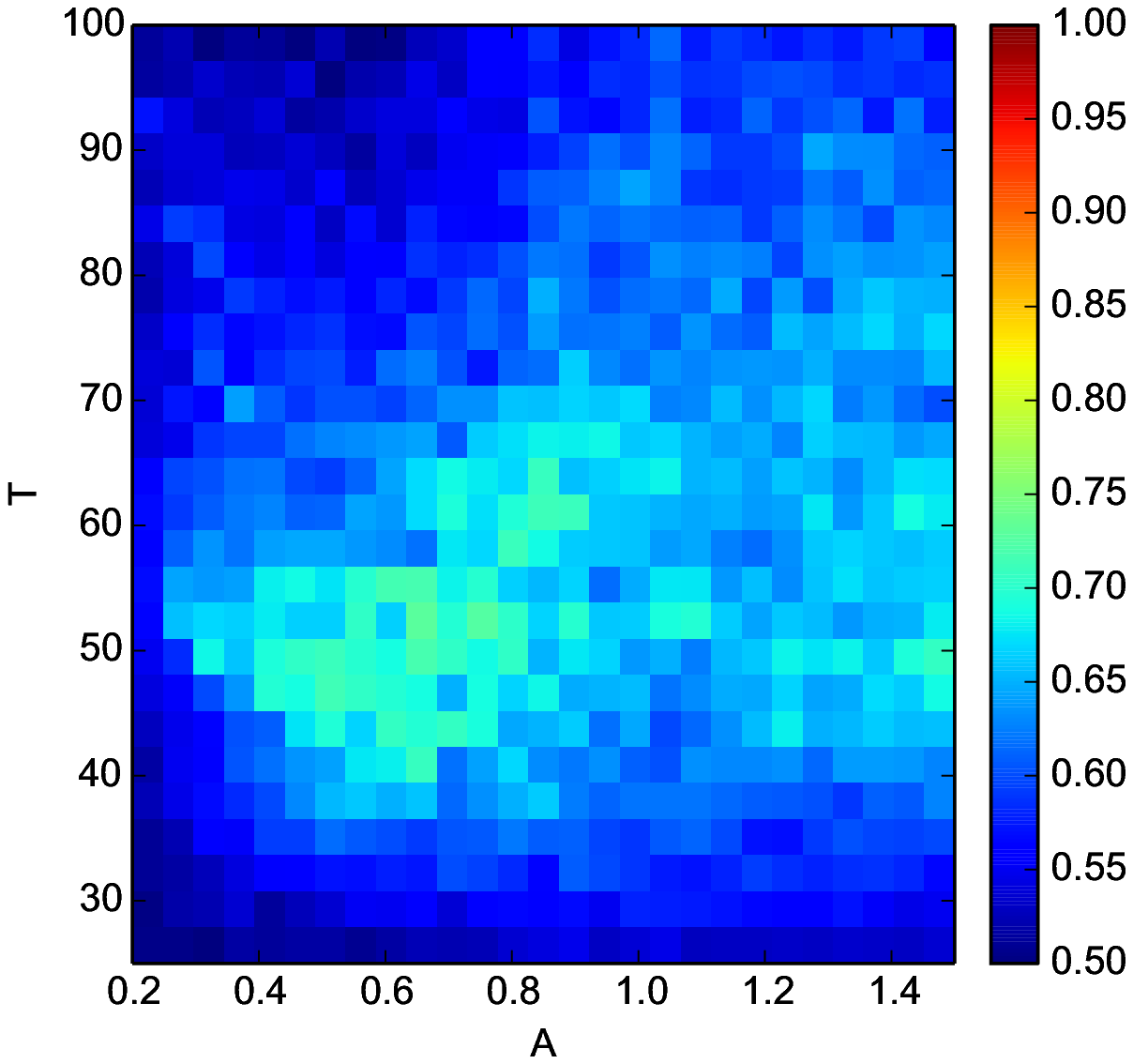} %Parity_P5.eps}
\caption{
{\bf Global tuning of network parameters.} Success probability ($P$) for the three parity functions ($P_3$, $P_4$, $P_5$, left to right) as the period $T$ of the input binary signal and the amplitude $A$ of the oscillator drive are varied globally for the whole network.
}
\label{fig:bandwidth}
\end{figure*}

Fig \ref{fig:bandwidth} shows variations in the success probability for the parity functions, as the global parameters $A$ (amplitude of oscillator drive) and $T$ (period of input binary signal) are varied.
Each network is trained and operated independently as described in section \ref{tasks}.
It is observed that the region of global parameter space where the performance of the network is good is reasonably large, indicating that the precise matching of the network to a given signal (especially with respect to $T$) is not required.
 
\begin{figure*}[ht]\centering % Using \begin{figure*} makes the figure take up the entire width of the page
\includegraphics[width=0.3\textwidth]{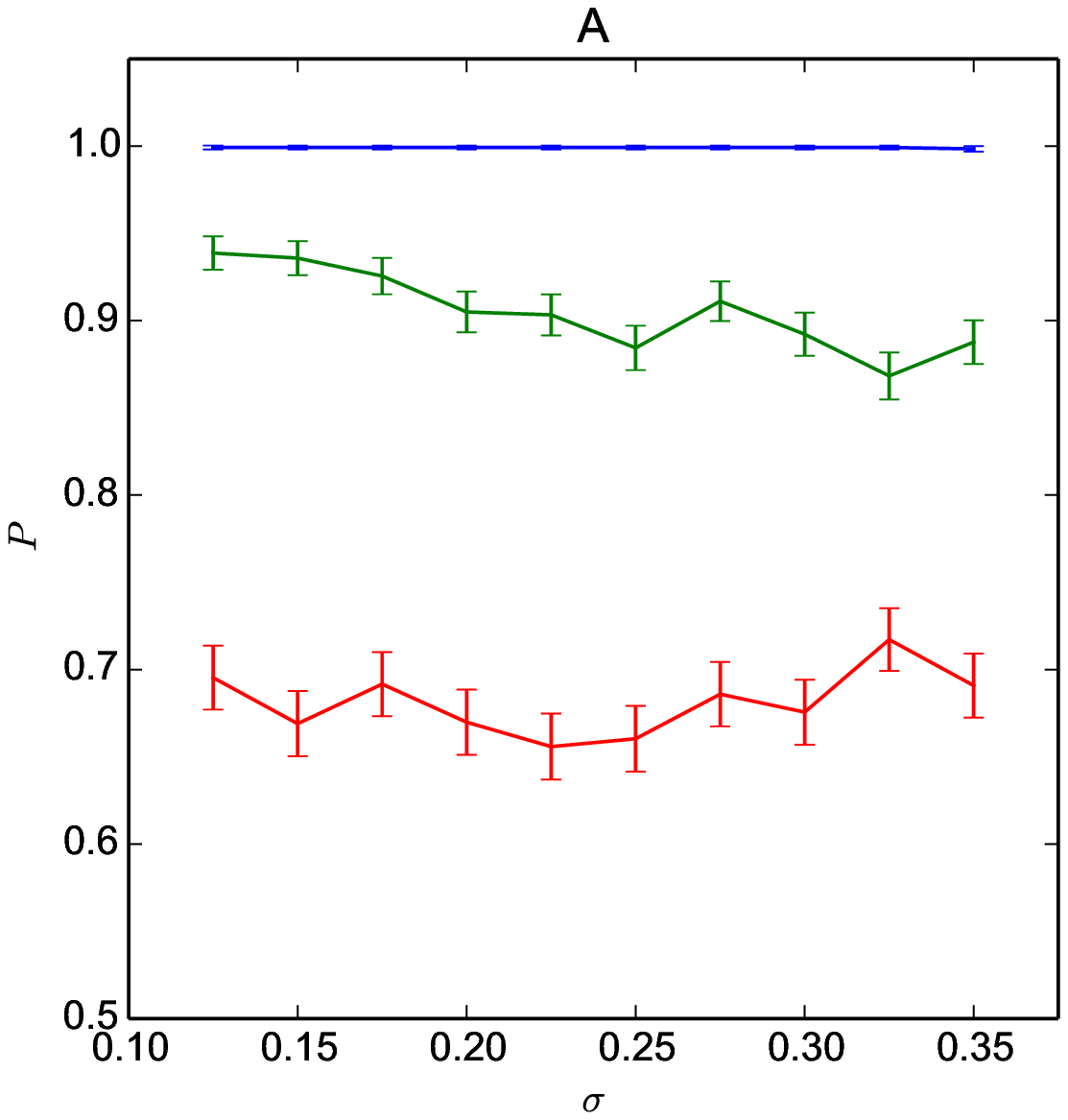} %A.eps}
\includegraphics[width=0.3\textwidth]{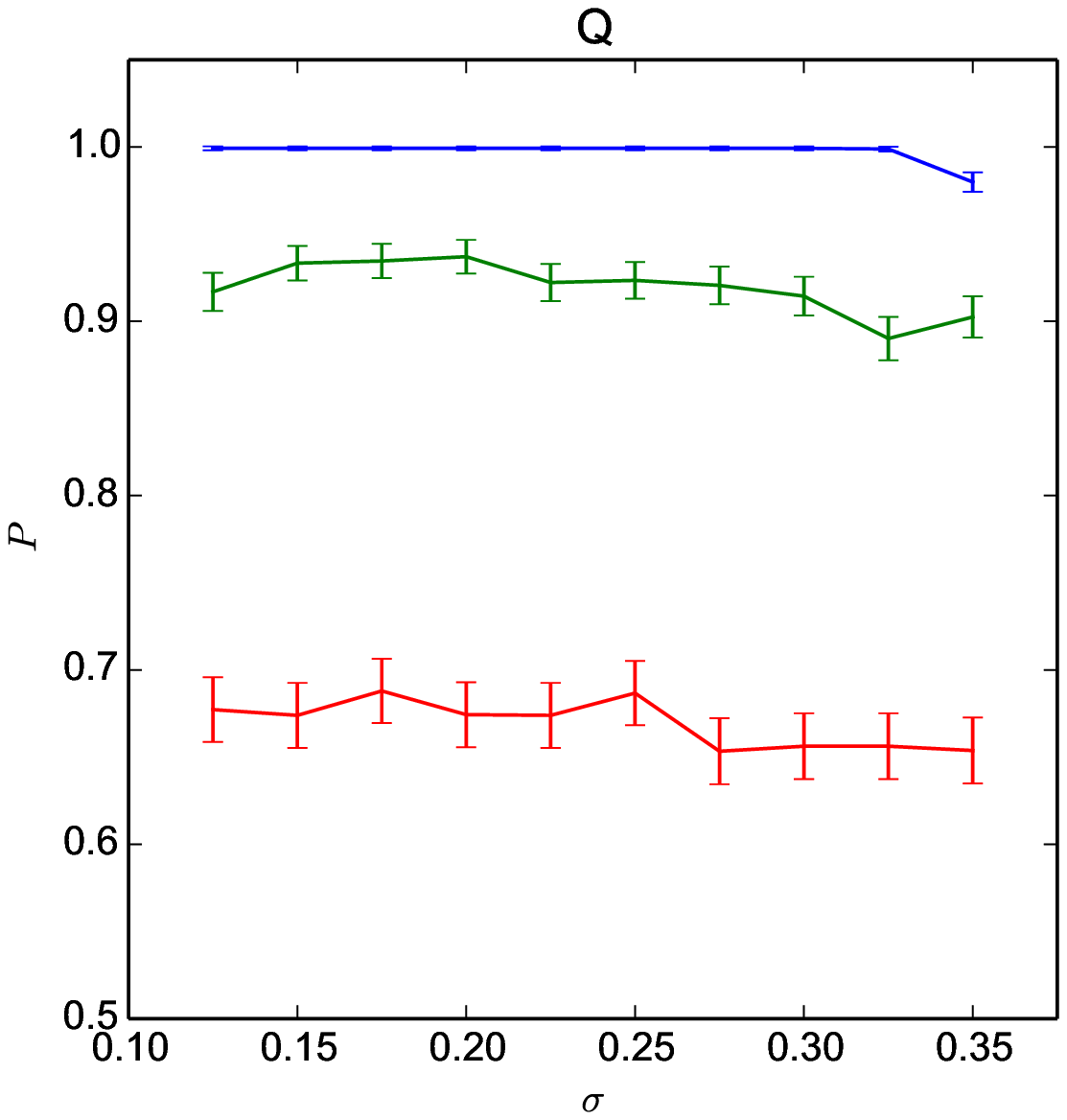} %Q.eps}
\includegraphics[width=0.3\textwidth]{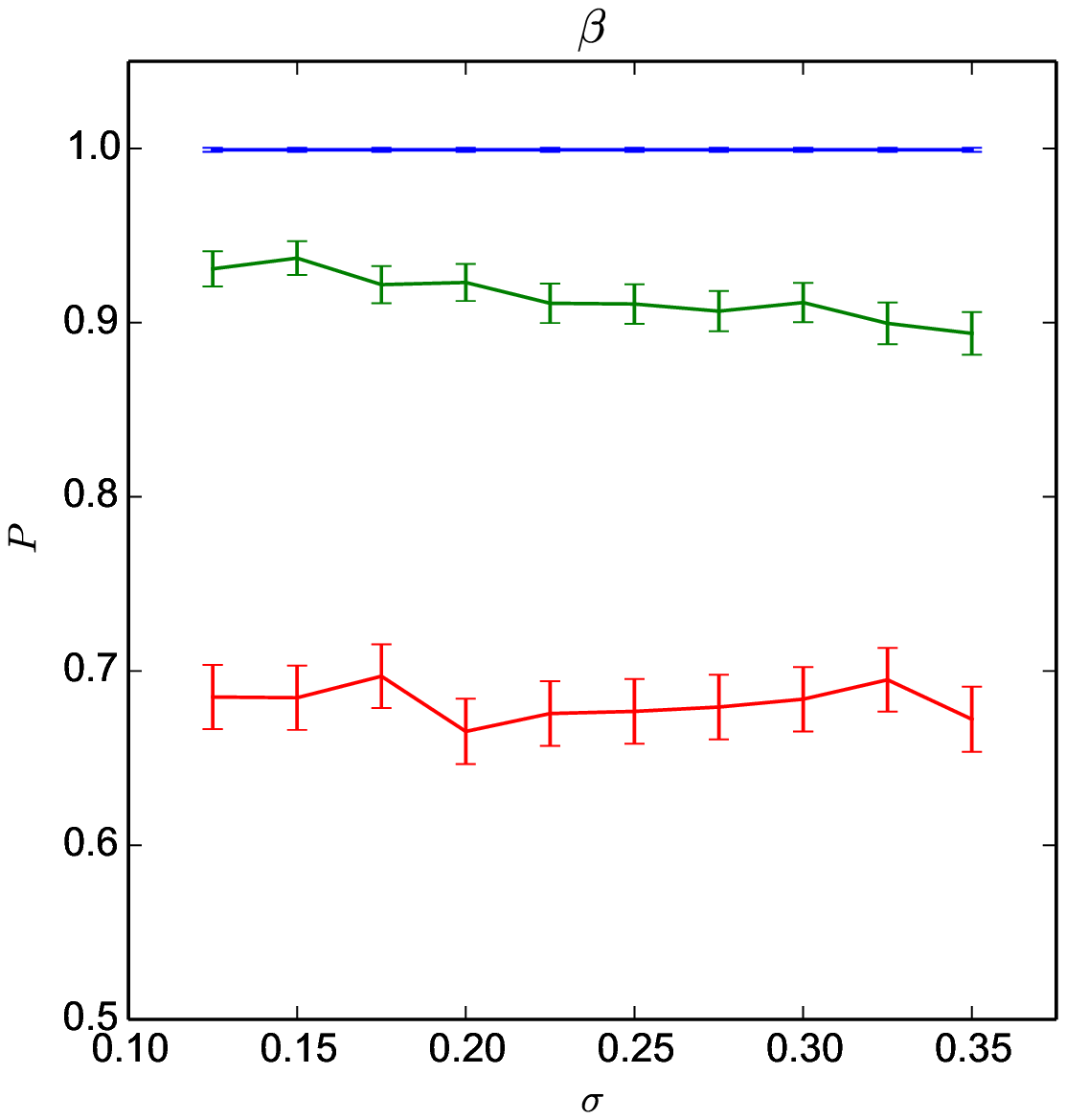} %beta.eps}
\includegraphics[width=0.3\textwidth]{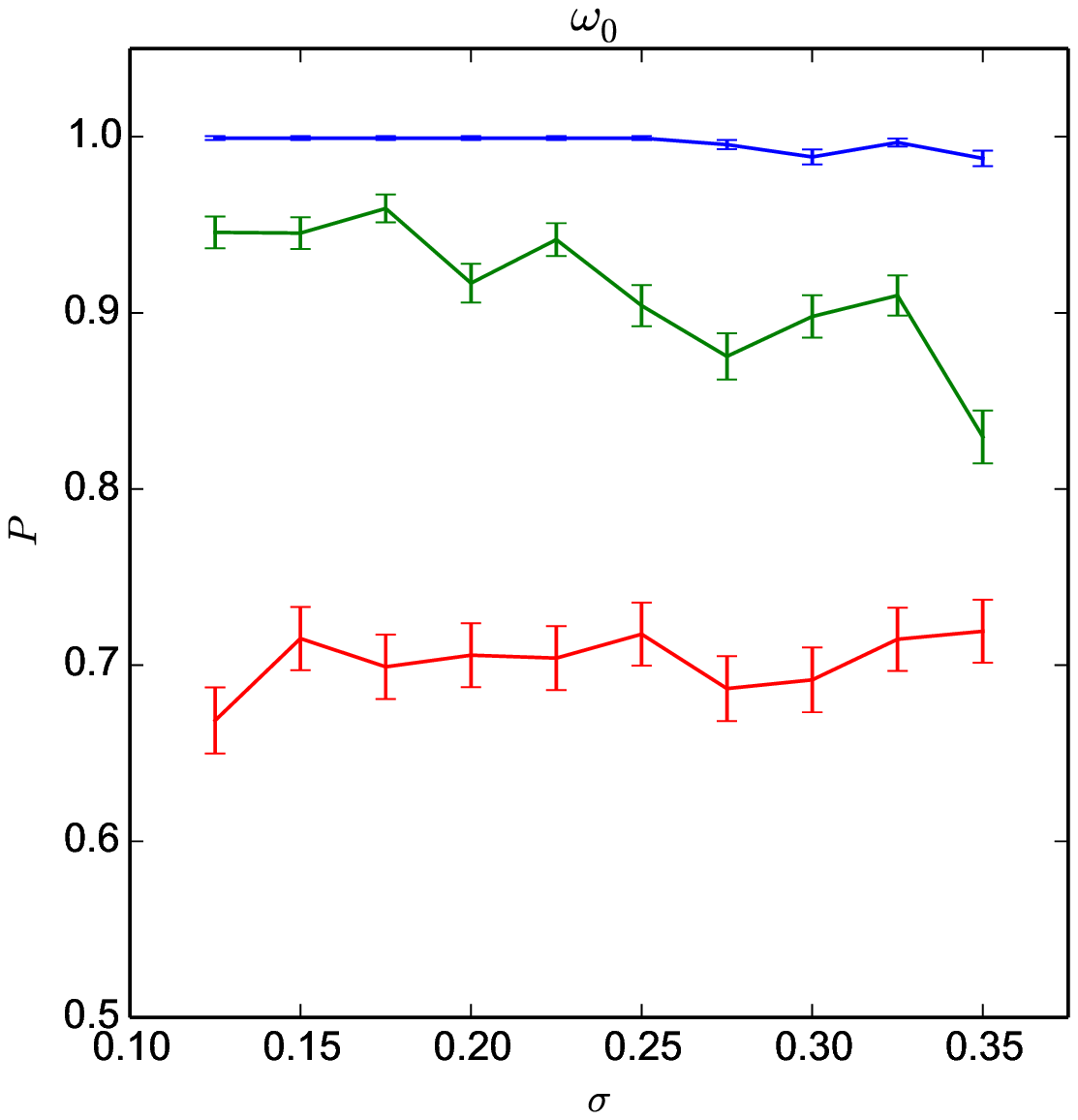} %w0.eps}
\includegraphics[width=0.3\textwidth]{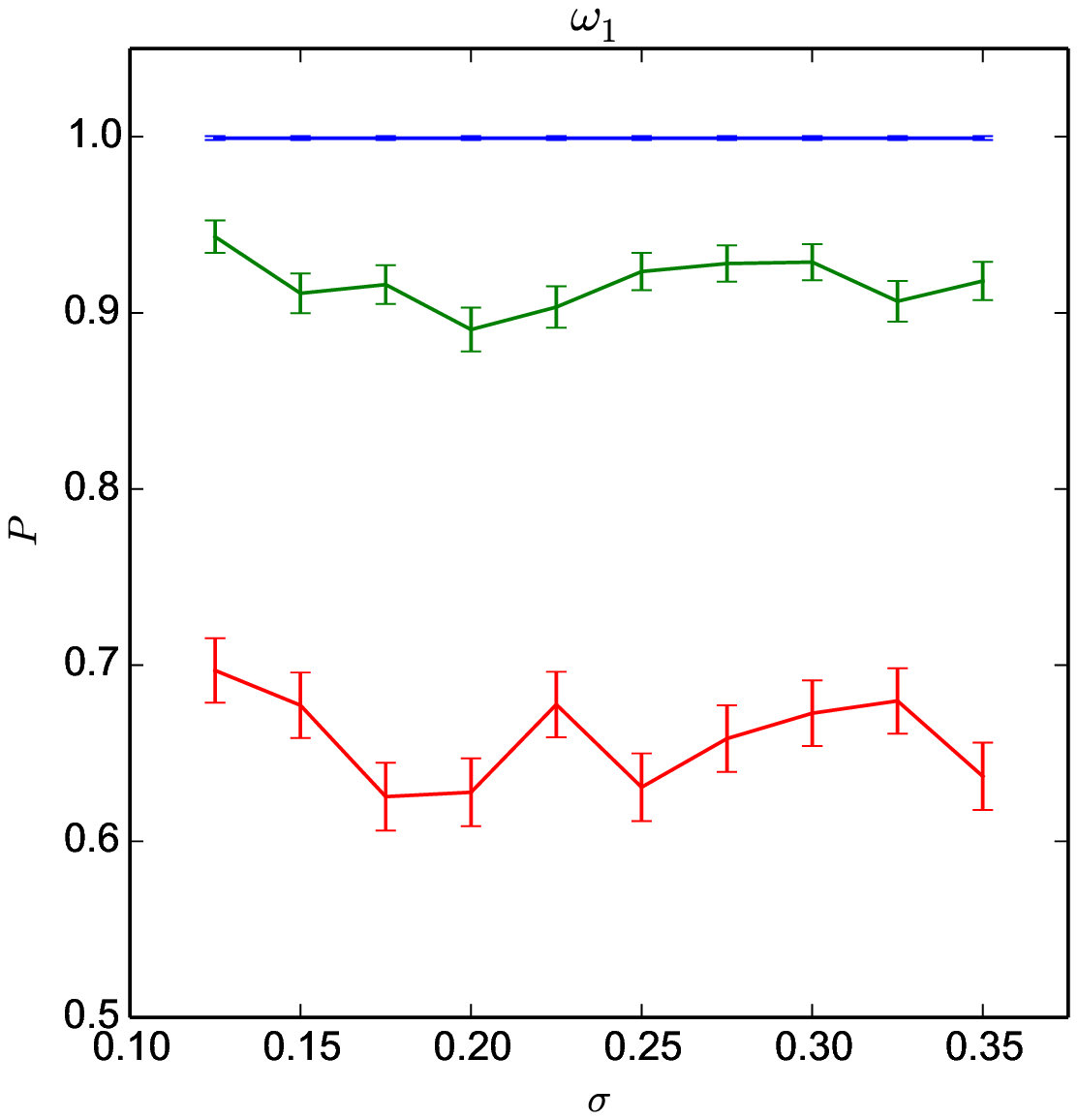} %w1.eps}
\includegraphics[width=0.3\textwidth]{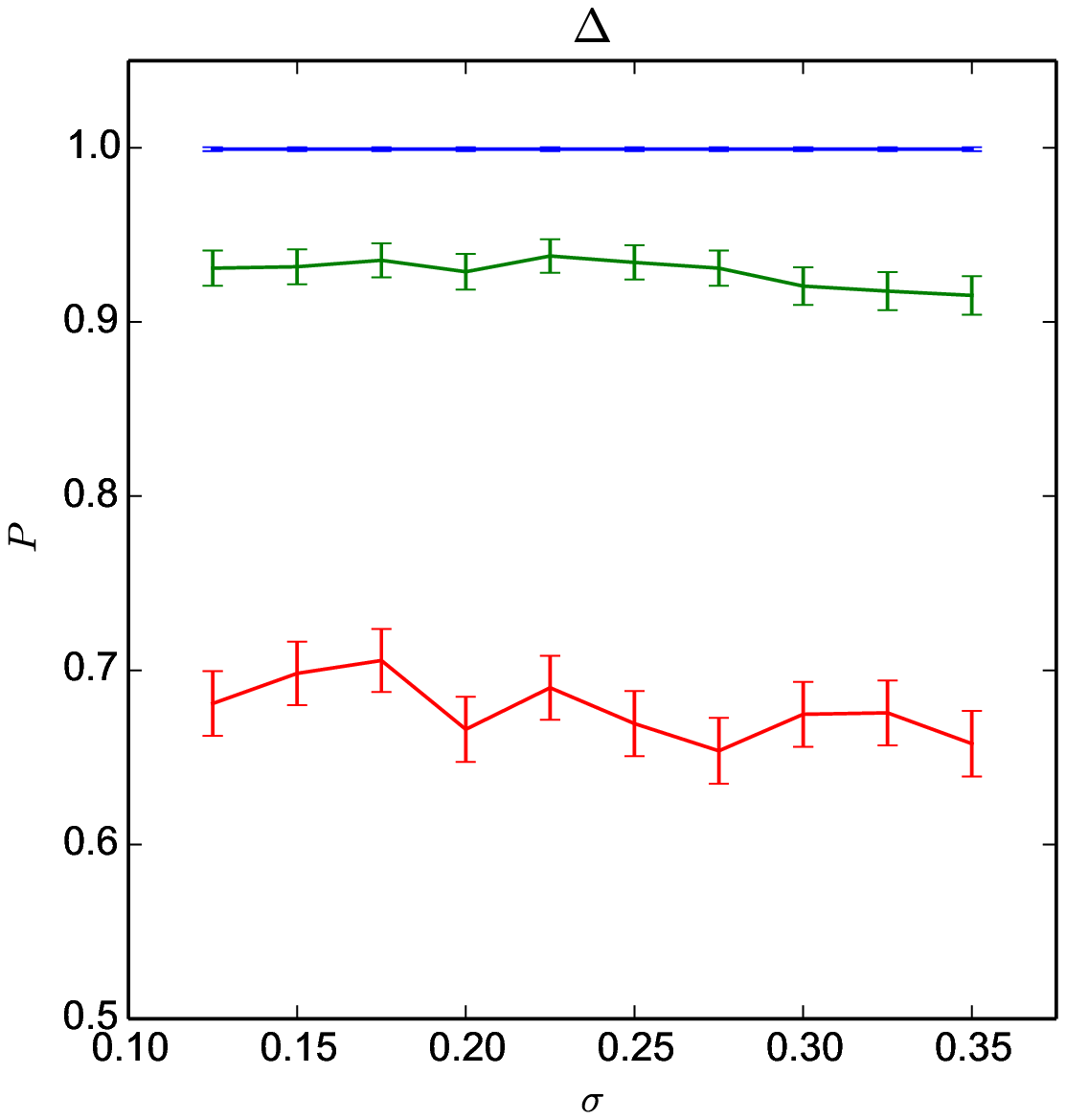} %dphi.eps}
\caption{
{\bf Robustness of the parity benchmark for pre-training variations.} Variations in the success probability ($P$) for the three parity functions (blue: $P_3$, green: $P_4$ and red: $P_5$) as the relative variation $\sigma$ is increased, for perturbations introduced before the training of the network is performed. Error bars are computed at the 95\% confidence level.
}
\label{fig:robustness}
\end{figure*}

Similarly, Fig \ref{fig:robustness} shows the reduction in the success probability for the parity functions, when the parameters of the oscillators in the network are not all identical, but are rather taken to fluctuate randomly, for instance to simulate the effect of manufacturing tolerances, according to
\begin{equation}
\lambda_i \rightarrow \lambda_i(1 + \sigma z), \label{eq:perturbations}
\end{equation}
where $\lambda_i$ can be any of the parameters in the set $\{A$, $Q$, $\beta$, $\omega_0$, $\omega_1$, $\Delta\}$ for the $i^{\rm th}$ oscillator, $\sigma$ is the relative fluctuation level, and $z$ is a standard normal random variable (zero mean and unit variance).
Each perturbed network is trained and operated independently as described in section \ref{tasks}.
While the performances of the network do depend on the nominal value of the parameters set for the oscillators, these data demonstrate for the parity function that the network is quite robust when the parameters of individual oscillators are independently varied around these nominal values.
In particular, the oscillators can be significantly detuned in frequency (variations in $\omega_0$) with the network still performing well on the parity benchmark, thus indicating that the dynamics of the network that are useful for computations are not the result of a precisely tuned resonant coupling of the oscillators.

\begin{figure*}[ht]\centering % Using \begin{figure*} makes the figure take up the entire width of the page
\includegraphics[width=0.3\textwidth]{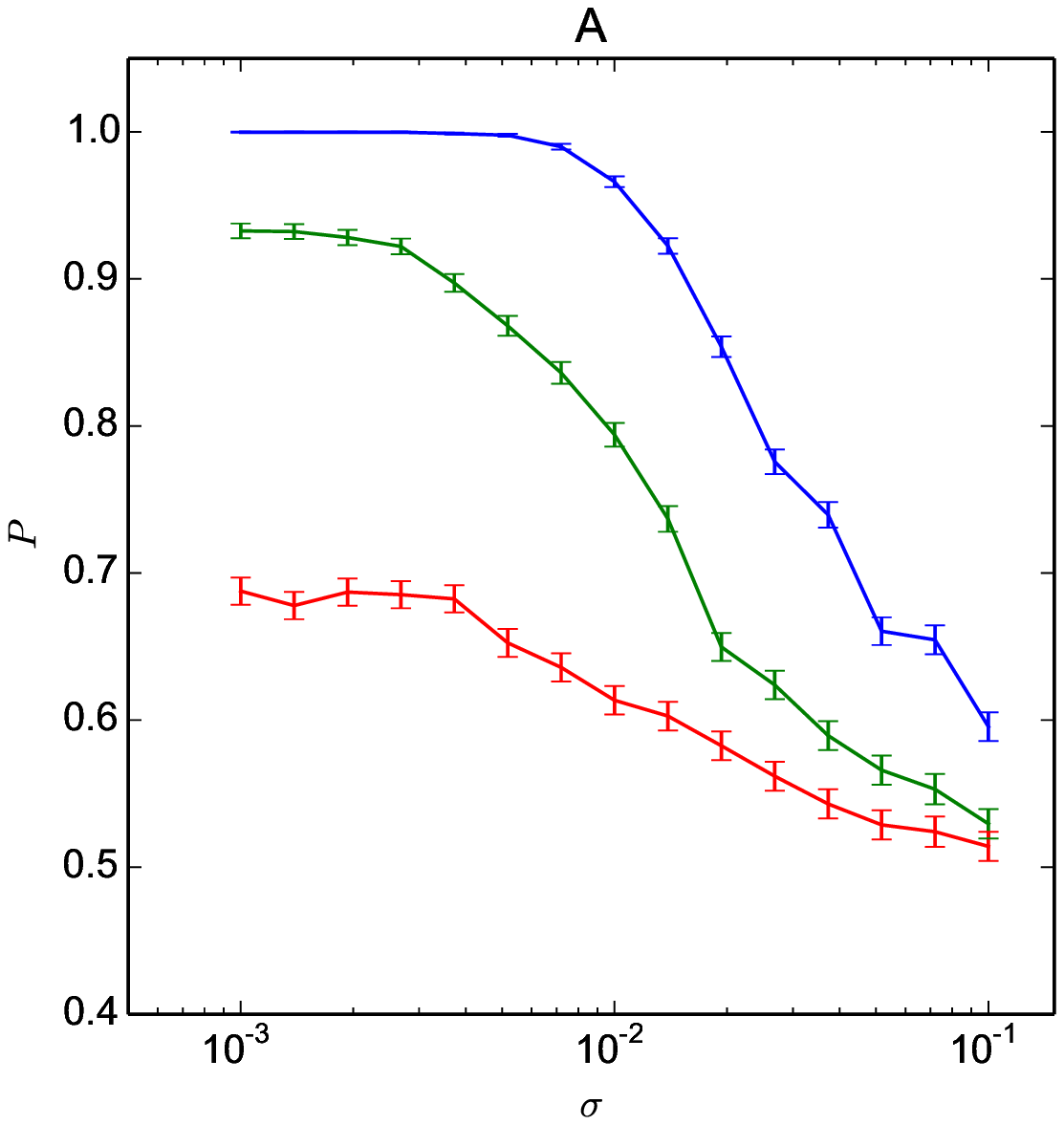} %PostA.eps}
\includegraphics[width=0.3\textwidth]{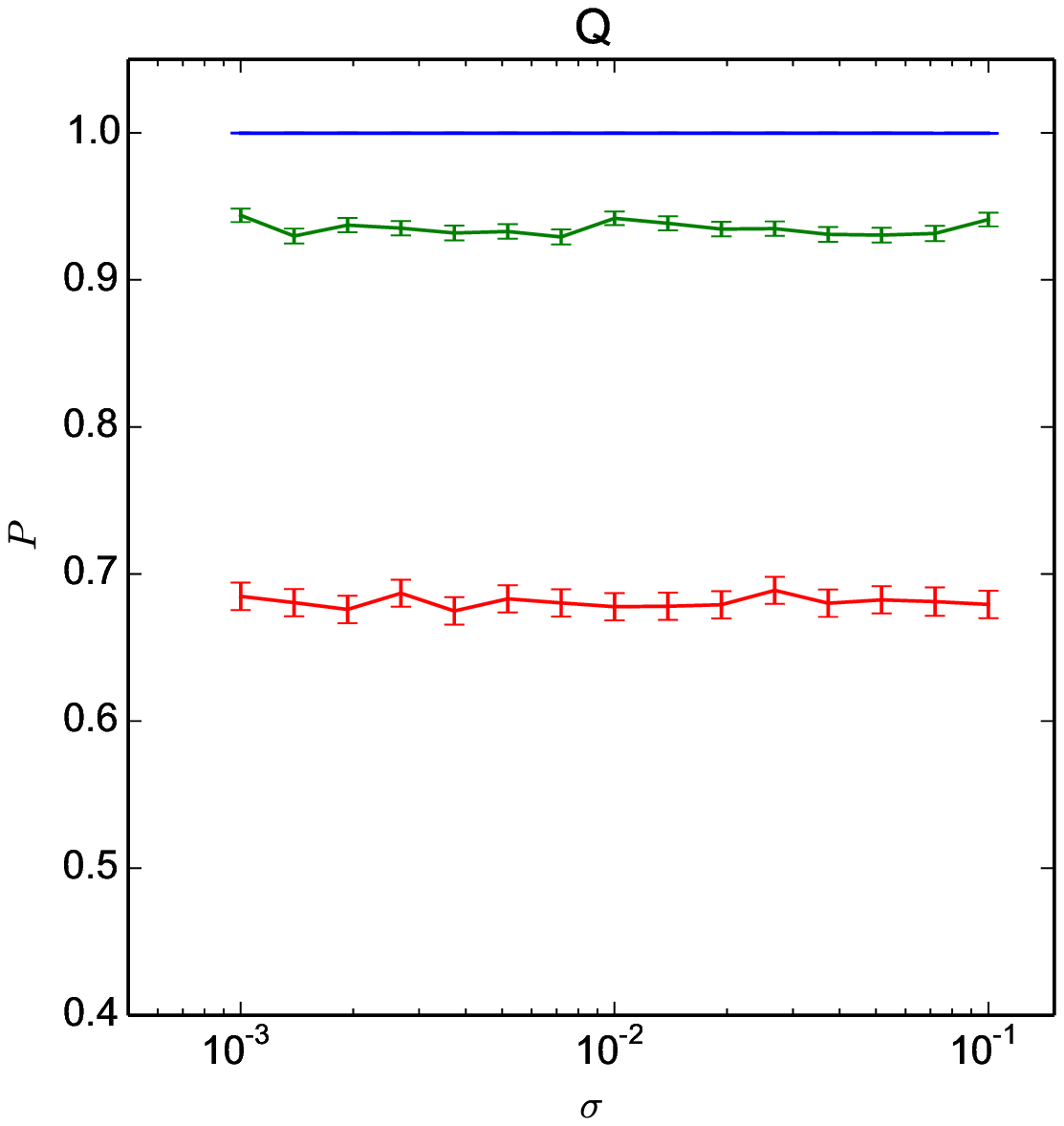} %PostQ.eps}
\includegraphics[width=0.3\textwidth]{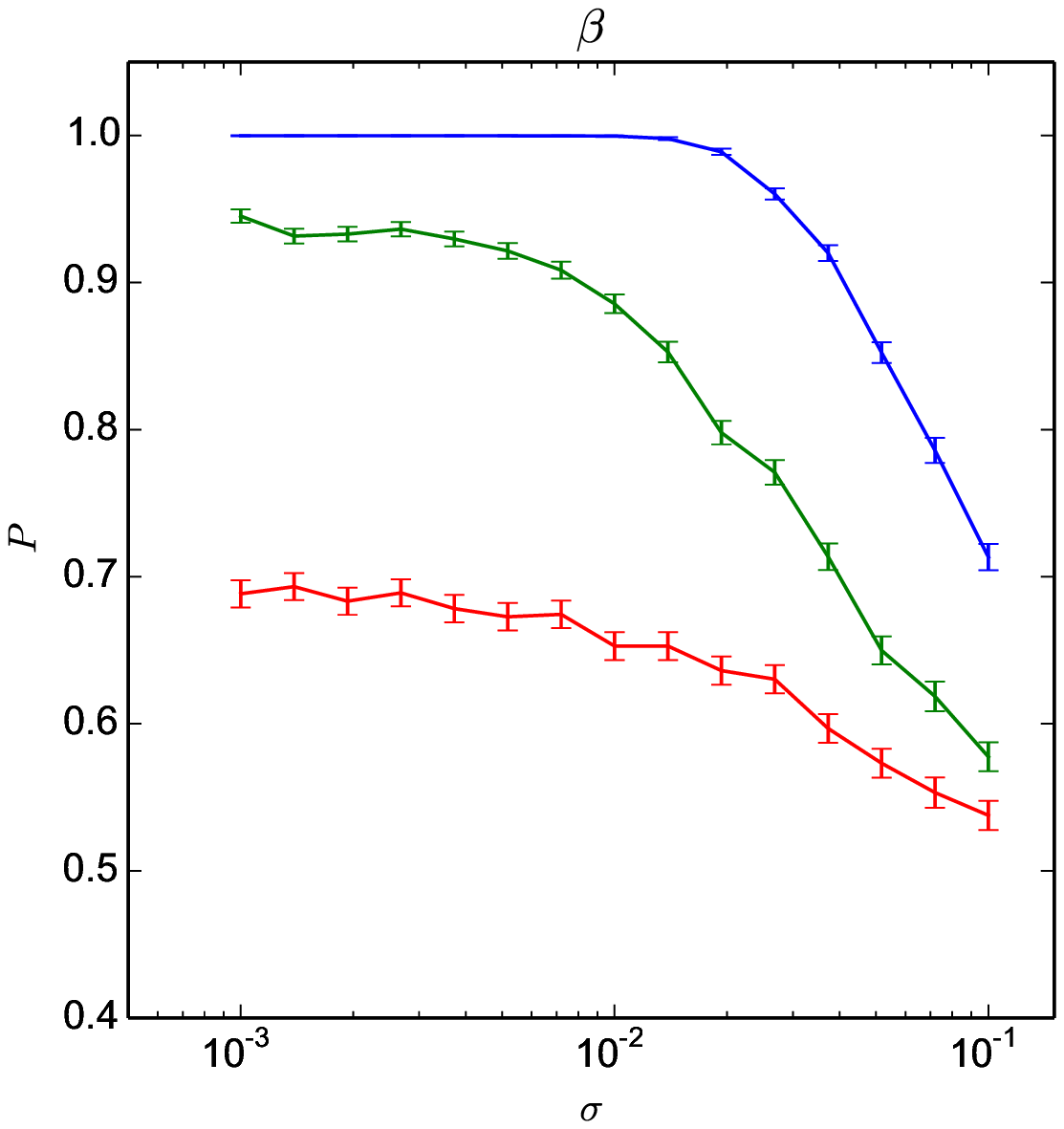} %PostBeta.eps}
\includegraphics[width=0.3\textwidth]{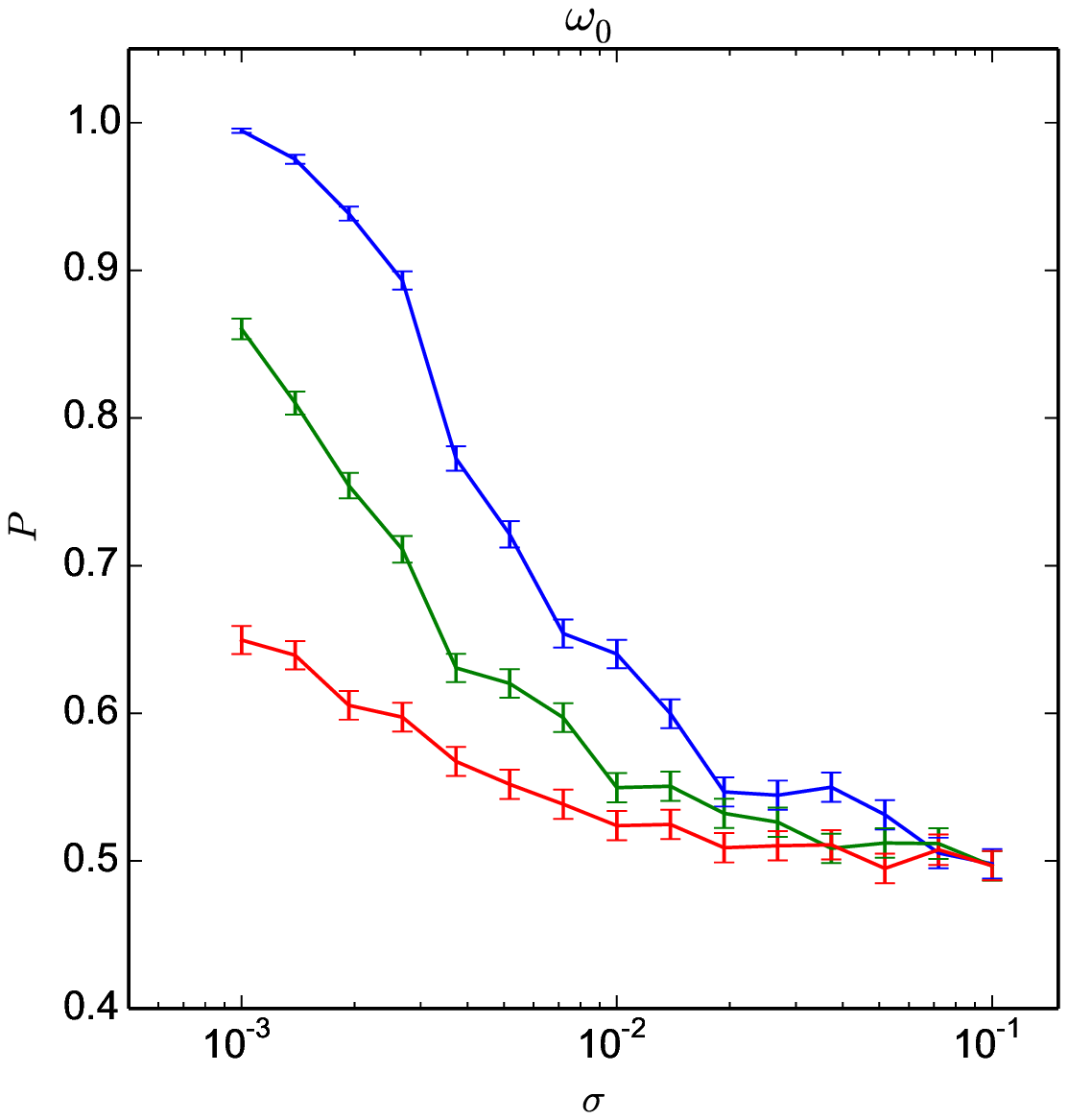} %Postw0.eps}
\includegraphics[width=0.3\textwidth]{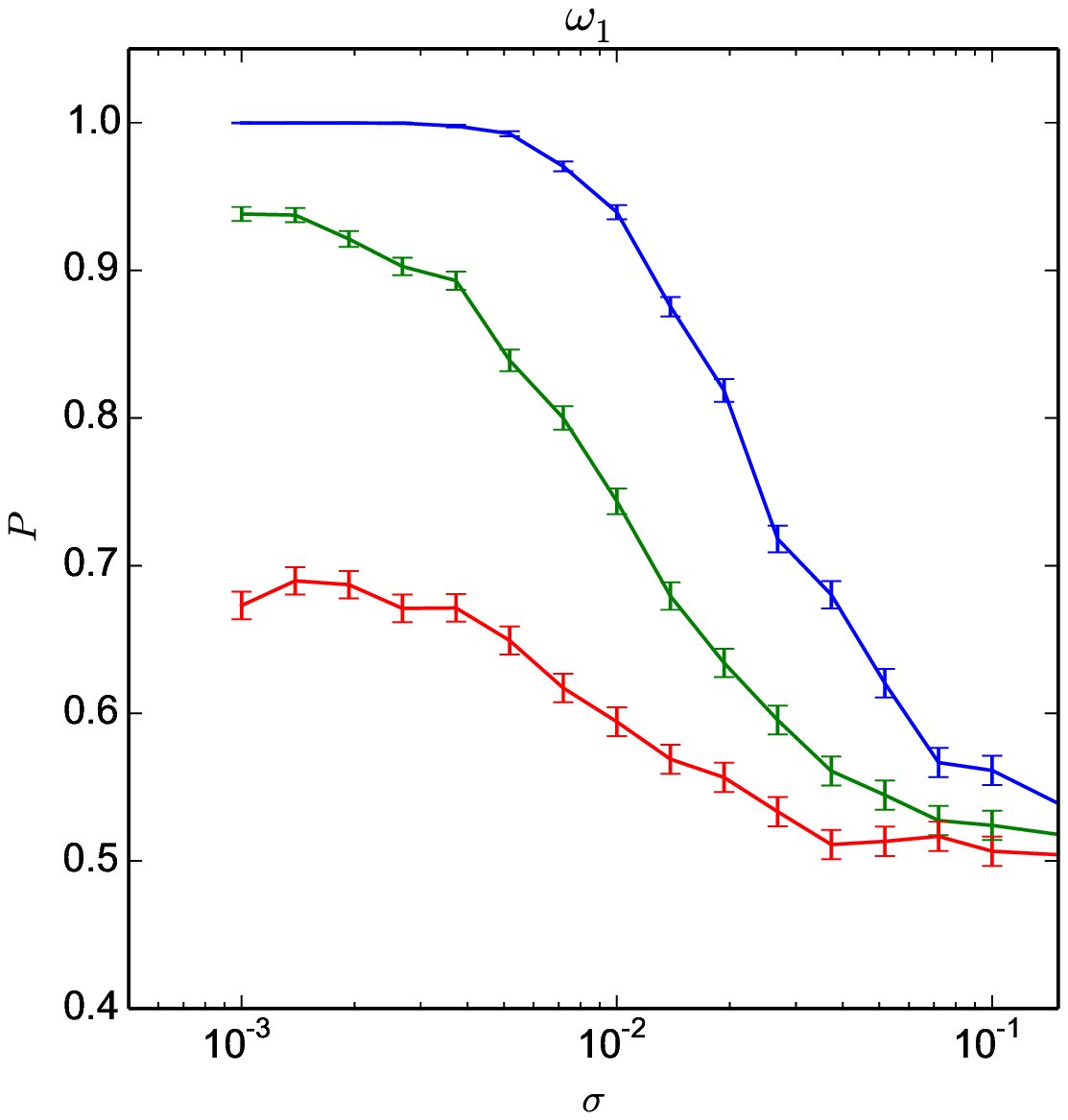} %Postw1.eps}
\includegraphics[width=0.3\textwidth]{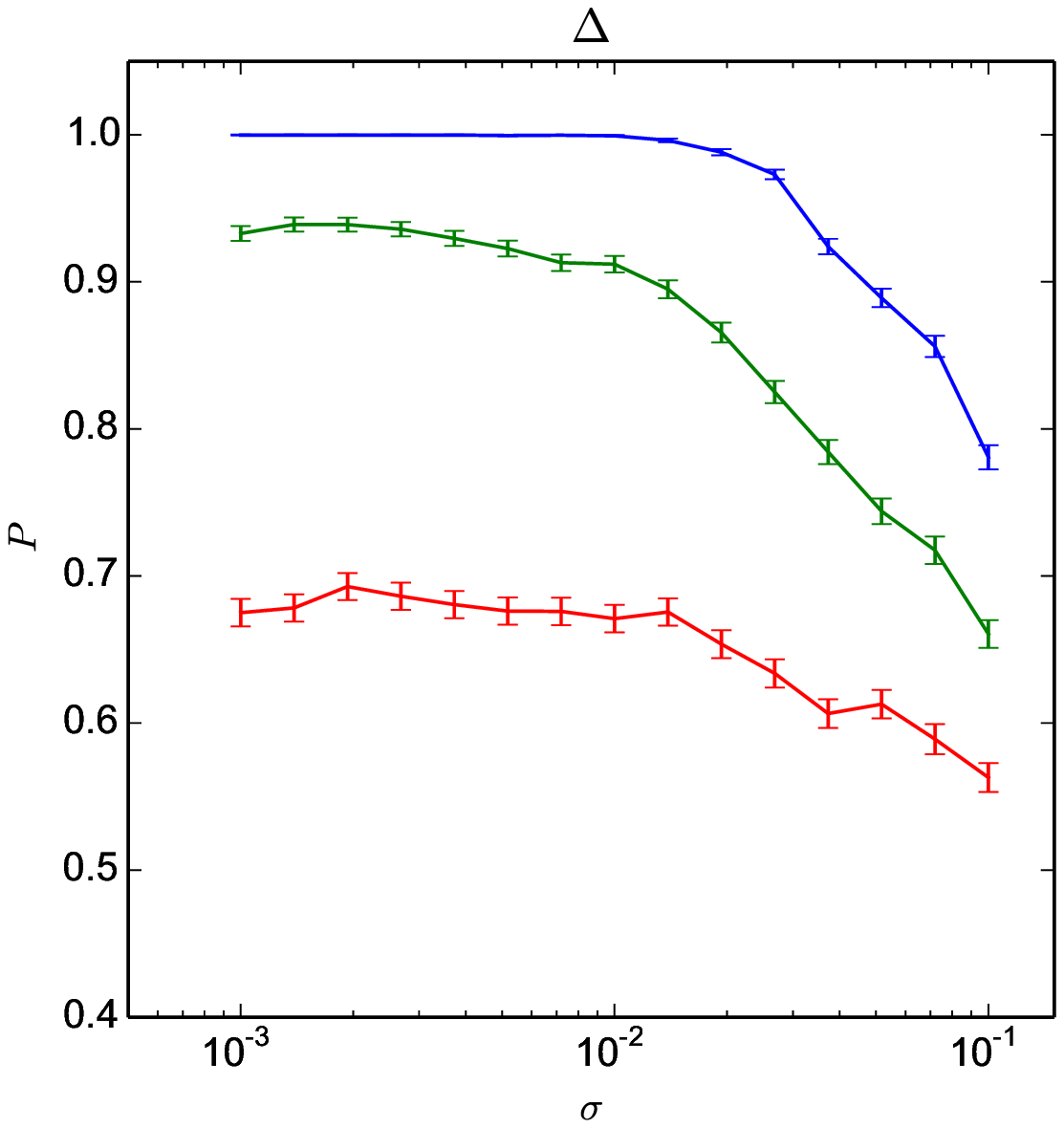} %Postdphi.eps}
\caption{
{\bf Robustness of the parity benchmark for post-training variations.} Variations in the success probability ($P$) for the three parity functions (blue: $P_3$, green: $P_4$ and red: $P_5$) as the relative variation $\sigma$ is increased, for perturbations introduced  after the training of the network is performed. Error bars are computed at the 95\% confidence level.
}
\label{fig:robustnessPost}
\end{figure*}

On the other hand, Fig \ref{fig:robustnessPost} shows how the success probability for the parity functions is reduced when perturbations described by Eq \ref{eq:perturbations} are introduced just after the training of the network has been completed.
This situation corresponds to oscillator parameters that are drifting over time.
It can be observed that the success probability is almost independent of variations in the quality factor ($Q$), for relative variations as high as 10\%.
For the non-linear parameter ($\beta$) and the coupling strength ($\Delta$), it is not reduced significantly for relative variations up to 1\%, but drops rapidly for large variations.
Finally, the success probability seems to degrade continuously with the magnitude of the relative variations for the harmonic drive amplitude ($A$), the coupling strength ($\omega_1$) and the oscillator natural frequency ($\omega_0$).
These observations are compatible with the hypothesis that computational capabilities depend strongly on the network being operated at a precise point with respect to its dynamics (as determined mostly by $A$, $\omega_1$ and $\omega_0$), presumably where the motion of the oscillators is complex, but not chaotic.
They also indicate that in actual physical devices, requirements on the stabilization of most parameters should be relatively mild ($\sim 1\%$), except for the parameters $A$, $\omega_1$ and $\omega_0$ which will have to be stable at the 0.1\% level.

\section{Conclusion}
Non-linear mechanical oscillators arranged in a network using linear couplings can be used to perform complex computations, as demonstrated in this communication by numerical simulations of a single instance of a network that performs well on two widely different benchmark tasks (computation of parity functions, and classification of spoken words).
The computational capacities of this network are preserved when the global network parameters are tuned to different values in a relatively large parameter space, when large (over 10\%) random variations are introduced in the oscillator parameters before the network training phase, and when significant variations (0.1\% to 1\%, depending on the parameter) in the oscillator parameters are introduced after the training phase has been completed with a network of oscillators having nominal parameter values.

\revision{These results show the existence of a new class of computing devices based on networks of coupled non-linear mechanical oscillators.
We have described one example of such device which, as mentioned above, performs robustly on two difficult computing tasks.
It is remarkable that this device has a low complexity (400 oscillators) relative to modern microelectronic components.
As explained in the introduction, this leads to the possibility of creating small and energy-efficient devices that are highly relevant technologically.

The device that was simulated in this work was constructed randomly from a small set of rules that are described in section \ref{tasks}.
These rules were only minimally tuned in order to obtain good performances on the benchmark tasks.
It was observed, for instance, that smaller networks did not perform as well on the spoken words benchmark.
It is a general characteristic of the reservoir computing approach that the details of the dynamical systems that are used for computing are not directly relevant.
However, it is likely that devices that differ from the one studied in this work might perform better, on a broader set of computing tasks.
Studies related to the optimization of parameters of mechanical oscillator networks, including the number of oscillators, their linear and non-linear characteristics, as well as the way they are interconnected, for instance, will be the subject of future communications.
}

As the mechanical oscillators can be directly coupled to forces produced by the environment of the network (accelerations, sound pressure, etc.), devices having both sensing and computing capabilities can be envisioned.
The devices, fabricated using MEMS technologies, for instance, could be very compact and energy-efficient, and compete with state-of-the-art sensors and microelectronic devices for distributed sensing or robot control.

\section*{Acknowledgments} % The \section*{} command stops section numbering
This work was supported by the Natural Sciences and Engineering Research Council of Canada (grant RGPIN-2015-05215) and by the Fonds de recherche nature et technologies du Québec (grant 2016-NC-189891).

\nolinenumbers

% Either type in your references using
% \begin{thebibliography}{}
% Text
% \end{thebibliography}
%
% or
%
% Compile your BiBTeX database using our plos2015.bst
% style file and paste the contents of your .bbl file
% here.
% 

%\bibliography{BiblioOscNetPaper}

\end{document}